\documentclass{article}
\usepackage[a4paper, total={6in, 8in}]{geometry}

\usepackage{listings}
\usepackage{amsmath,amssymb}
\usepackage{graphicx}
\usepackage{latexsym}
\usepackage{xcolor}
\usepackage{url}
\usepackage[bookmarks=true,bookmarksopen=true]{hyperref}
\usepackage[T1]{fontenc} 
\usepackage[utf8]{inputenc}
\usepackage{subfigure}
\usepackage{bm}
\usepackage{booktabs}
\usepackage{authblk}
\usepackage{orcidlink}
\usepackage{hyperref}

\usepackage{natbib}
\bibpunct[, ]{(}{)}{;}{a}{}{,}

\providecommand{\keywords}[1]
{
  \small	\noindent
  \textbf{Keywords:} #1
}

\title{Spatio-Temporal Analysis of Public Transportation Undercrowding: Leveraging APC Data for a Comprehensive Evaluation of Usage Rates}

\author[1]{Arianna Burzacchi$^*$}
\author[2]{Valeria Maria Urbano$^*$}
\author[2]{Marika Arena}
\author[2]{Giovanni Azzone}
\author[1]{Secchi Piercesare}
\author[1]{Simone Vantini}

\affil[1]{MOX Laboratory, Department of Mathematics, Politecnico di Milano, 20133, Milano, Italy}
\affil[2]{Department of Management, Economics and Industrial Engineering, Politecnico di Milano - Via Lambruschini 4/B, 20156, Milano, Italy}

\date{}

\begin{document}

\maketitle
\def\thefootnote{*}\footnotetext{Corresponding authors. Emails: \{arianna.burzacchi, valeriamaria.urbano\}@polimi.it}

\def\thefootnote{\arabic{footnote}}

\abstract{
The analysis of the transportation usage rate provides opportunities for evaluating the efficacy of the transportation service offered by proposing an indicator that integrates actual demand and capacity. This study aims to develop a methodology for analyzing the occupancy rate from large-scale datasets to identify gaps between supply and demand in public transportation. Leveraging the spatio-temporal granularity of data from Automatic People Counting (APC) and relying on the Generalized Linear Mixed Effects Model and the Generalized Mixed-Effect Random Forest, in this study we propose a methodology for analyzing factors determining undercrowding. The results of the model are examined at both the segment and ride levels. Initially, the analysis focuses on identifying segments more likely associated with undercrowding, understanding factors influencing the probability of undercrowding, and exploring their relationships. Subsequently, the analysis extends to the temporal distribution of undercrowding, encompassing its impact on the entire journey. The proposed methodology is applied to analyze APC data, provided by the company responsible for public transport management in Milan, on a radial route of the surface transportation network. 
}


\newpage
\section{INTRODUCTION}\label{sec:intro}

In public transportation, monitoring the transportation usage rate provides opportunities for evaluating the quality of the transportation service offered by including the actual demand and the capacity of vehicles in one single indicator. Usage rate has been quantitatively assessed by using occupancy rate, or load factor, which is defined as the ratio between the actual number of passengers inside a transportation vehicle and the capacity of the vehicle (either measured considering the number of seats or the sum of seating and standing places) \citep{whelan2009investigation, jara2003towards}. The evaluation of utilization rates and the identification of critical situations in transportation usage rates is particularly relevant because it can be approached from two divergent perspectives. From the standpoint of transport operators, a high load factor is desirable for sustainability and efficiency, translating to lower emissions per passenger and improved resource allocation. Conversely, from the user's viewpoint, the load factor serves as a pivotal indicator of service quality. A high load factor, while advantageous for operators, can lead to crowded conditions, causing discomfort for passengers \citep{de2013perceived, dell2011quality}. Consequently, monitoring the load factor of public transport modes is crucial to facilitate a timely response plan from public transport operators.

During the last decades, the analysis of the occupancy rate of vehicles in public transportation has been based primarily on data collected through surveys. Surveys data are collected on a yearly base at most, and allow for the estimate of the average travel habits of individuals, hence not providing opportunities for investigating variability in space and time. With the emergence of intelligent transport systems and the development of more cost-efficient data collection, new data sources, including automatic vehicle location, automated fare collection systems, automated people counting systems, or weight sensors, became available \citep{zou2022passenger,komatsu2021passenger}. Compared to traditional sources of data, novel data sources provide continually transmitted data that allows to analyze travel behavior with spatial and temporal granularity. However, big data sources present also new challenges associated with the identification and extraction of meaningful insights from large datasets \citep{mandelzys2010identifying, berkow2009beyond}. 

In response to the increasing demand for methodologies tailored to large-scale datasets, as emphasized by \citet{hellinga2010automatically, mandelzys2010identifying}, this study introduces a two-step methodology designed to support service planning in urban public transportation. Leveraging the spatio-temporal granularity of Automatic People Counting (APC) data, our study aims to develop a methodology for efficiently analyzing the occupancy rate from large-scale datasets. The methodology supports the detection of critical situations stemming from imbalances between supply and demand in public transportation through the analysis of load factors. To this end, we integrated APC data from two additional sources: weather information and specific events data (such as strikes, national or local holidays, etc.). The integration of these data sources enhances the capacity to identify gaps between supply and demand by providing a more comprehensive understanding of the underlying causes behind critical situations.

In the first step, we implemented two prediction models (using statistical and machine learning approaches) to predict the probability of overcrowding and undercrowding situations at a single segment level. These situations serve as indicators of potential service quality issues. A comparative analysis was conducted to evaluate the accuracy of various prediction models. Results of the model provided insights on segments more likely associated with undercrowded, factors significantly influencing the probability of undercrowding, and their relationship.

The second step aims at identifying undercrowding at the ride level. In this step, we implemented a methodology for identifying the spread of undercrowding or overcrowding throughout the entire journey and scrutinizing the distribution of critical situations by time slot, day type, and week. By discerning specific patterns over time, transportation planners can pinpoint potential interventions and improvements to rectify the gap between supply and demand.

Through the proposed two-step approach, our study aims to contribute to the advancement of efficient and effective service planning activities in urban public transportation using APC data. Our empirical analysis primarily delves into the investigation of undercrowding, but it is crucial to emphasize that the methodology we have developed can be readily adapted to evaluate the level of comfort in transportation services by examining overcrowding situations. The versatility of this methodology allows for a comprehensive examination of both undercrowded and overcrowded scenarios, contributing to a more holistic understanding of the dynamics and quality of transportation services.

The paper is structured as follows. In Section \ref{sec:SOTA}, we offer an overview of the state of the art concerning the utilization of APC data. Section \ref{sec:method_data} introduces the data sources and methodologies employed in this study. It details the pre-processing methods applied to the data and outlines the model developed to estimate the probability of undercrowding or overcrowding, taking temporal factors, weather, and specific event data into account. In Section \ref{sec:results}, the results are provided and discussed. Finally, in Section \ref{sec:conclusions} conclusions are presented, including practical implications of the study and further areas of research. 

\section{STATE OF THE ART} \label{sec:SOTA}

APC systems are passenger volume technologies that can utilize either direct or indirect measurements \citep{olivo2019empirical}. Indirect measurement methods include weight-based and mobile device-based systems. The former collects information on the total weight of passengers, providing the total number of passengers without differentiating boarding or alighting counts. Mobile-based systems, on the other hand, may yield less accurate estimates due to varying device penetration rates, particularly among different age groups.

Direct measurement methods accurately recognize the number of people boarding or alighting from vehicles. Three main technologies are employed for direct measurement: old mats technologies, infrared technologies, and imaging technologies. Infrared technologies utilize light beams to measure passenger volumes, with the sequence of beams indicating the direction of passenger movements (boarding or alighting). A considerable body of literature has focused on analyzing the strengths and weaknesses of these technologies and evaluating the accuracy of their estimates \cite{olivo2019empirical,siebert2020validation,moser2019methodology,saavedra2011automated,kovacs2008automatic,barabino2014offline}.

Researchers explored the opportunities provided by the analysis of APC data,  developing new methodologies that support decision-making for public transport operators. Some studies have integrated APC data with timetabling information to assess the impact of passenger volume on dwell time. Understanding the relationship between passenger flows and dwell time is crucial for identifying causes of timetable instability and network inefficiencies, which can then inform timetable planning \citep{buchmueller2008development, christoforou2020investigating, milkovits2008modeling}.

Performance monitoring and service planning have also been the focus of several studies using APC data. Some authors present methodologies for offering insights into system usage and passenger behavior \citep{wilson2008potential}. Other authors provided systematic approaches for understanding performance from various perspectives, such as passenger waiting time, bus bunching levels, and bus occupancy \citep{pi2018understanding}. Further methodologies have been proposed to identify situations where performance standards were not met and to determine factors contributing to quality issues \citep{mandelzys2010identifying, hellinga2010automatically}. These studies provide frameworks and methodologies for analyzing large volumes of data in an automatic way, hence extracting meaningful information on the entire transportation system. 

Several authors have addressed specific problems in service planning using APC data. For instance, some studies have analyzed APC data to understand transit vehicle delays and propose measures for transit priority \citep{hellinga2011estimating, yang2012estimating}. Others have tackled the frequency setting problem, aiming to optimize efficiency for operators and level of service for users \citep{hadas2012public}. Missed transit connections were addressed by providing a systems approach to quantify the impact of travel time reliability, schedule adherence, and schedule design \citep{mai2012simulating}. Additionally, bus schedule reliability has been studied using automated learning frameworks and robust optimization models \citep{khiari2016automated, baghoussi2018updating}. Passenger waiting time has been a focus of real-time bus holding control strategies aimed at minimizing total passenger waiting time \citep{mohamadamin2017real}.

More recently, forecasting methodologies have been applied to predict both travel time and occupancy rates using APC data. Travel time predictions have been estimated using regression models, as demonstrated by \citet{soroush2011predicting}. On the other hand, more sophisticated approaches have been explored, such as the development of Artificial Neural Network (ANN) models for predicting bus travel time, which has proven to yield accurate results \citep{arhin2020predictive}.

Forecasting the occupancy rate has garnered significant attention from researchers. In a study by \citet{kim2009comparative}, the goodness-of-fit of aggregate and disaggregate gravity modeling using APC data was compared to the disaggregate modeling approach, outperforming the aggregate model. Another approach involved supervised learning of a regression model, adopted by \citet{samaras2015prediction}, to predict passenger flow at single stops and routes, yielding promising results compared to baseline approaches.

In the last few years, machine-learning approaches have been extensively explored for such predictions. Decision tree methodologies have been combined with bagging and boosting techniques to predict ridership levels \citep{karnberger2020network}. Density-Based Spatial Clustering of Applications with Noise (DBSCAN) in conjunction with the Seasonal Autoregressive Integrated Moving Average (SARIMA) algorithm was utilized to analyze real-time passenger demand forecasts dynamically, endorsing the growth of dynamic bus management \citep{thiagarajan2021identification}. The same methodology (SARIMA) was implemented by authors to combine short-term prediction with real-time data \citep{hoppe2023improving}. Other scholars used the Kalman filter approach and support vector regression algorithm to predict short-term passenger flows \citep{wang2021atwostage}.

\section{METHODOLOGY AND DATA}\label{sec:method_data}
To develop the methodology we relied on data provided by Azienda Trasporti Milanesi (ATM), the company responsible for the management of public transport in Milan, which manages both the surface transport and the underground transport of the municipality, provided data from APC of vehicles moving on the surface of the city. The surface transportation network consists of $19$ tramway lines over $180$ kilometers of network, $134$ bus lines, and $4$ trolley bus lines covering about $1,500$ kilometers. By providing $25$ thousand rides every day, the surface transportation network meets the mobility needs of more than $400$ million passengers per year. 

The sample of data considered in this paper includes bus rides on a radial route linking a central point in town with a suburban area, $11,672$ from the center to the suburbs, and $11,643$ in the opposite direction. Although both routes were studied, the results of the empirical analysis will be presented only for the route from the city center to the suburbs. Vehicles on this route stop at $19$ different bus stops linking a terminal stop of a metro line leading to the city center to a peripheral area in the Milan suburbs. The dataset covers a period of 6 months, providing data on occupancy, boarding, and alighting passengers on the two routes for 26 weeks between the 6th of June to the 5th of December 2022. 

The dataset contains 45 attributes, that can be divided into 9 ride-related attributes and 36 station-related attributes. Table \ref{tab:raw_data} outlines the selected subset of station-related and ride-related information chosen for the analysis, representing a subset of the overall set of attributes. In terms of ride-related information, the ID of the ride is determined by a combination of five different variables, i.e. the Date, the Route, Table, Ride, and Direction. Additionally, details about the type of vehicle, the direction of travel, and the specific route path can be found in the data. In terms of station-related information, the dataset provides details about various aspects including the service status, the closest bus stops, and the number of passengers boarding and alighting at each station.

\begin{table}[htb]
    \centering
    \scriptsize
    \caption{Raw data description}
    \begin{tabular}{ccp{7cm}}

        \toprule
        \textbf{Type} & \textbf{Variable} & \textbf{Variable description}  \\
        \midrule

        Ride-related data & Date & Date of the ride \\
        & Route & Route number \\ 
        & Table & Table number  \\
        & Ride & Ride number  \\
        & Direction & Route direction with value 0 for Outbound direction, 1 for Inbound direction, and 2 for non-in-transit vehicles\\
        & Vehicle & Vehicle number \\
        & Vehicle type & Vehicle type with value 0 for Tram, 1 for Bus, and 2 for Missing value\\

        & Path & Route path number\\
        \midrule

        Station-related data & TimeStamp & Precise measurement time, including both date and time information  \\ 
        & Noise & Binary variable with value 1 for noisy data and 0 for non-noisy data \\
        & Status & Operating service status, from which to extract the information on operative status distinguishing between In-transit observations and Not-In-transit observations \\
        & Diverted route & Diverted route number \\
        &  Stop & Id of che closest stop  \\
        
        & APC Info type &  Category variable with value 0 for No information on passengers, 2 for counting of passengers, 7 for cumulative counting of passengers\\
        & INF1 & On-board passengers, available when APC Info type = 7 \\
        & INF2 & Boarding passengers, available when APC Info type = 2; Cumulative boarding passengers, available when APC Info Type = 7  \\
        & INF3 &  Alighting passengers, available when APC Info type = 2; Cumulative alighting passengers, available when APC Info Type = 7  \\
        
        \bottomrule

    \end{tabular}
    \label{tab:raw_data}
\end{table}

As a preliminary step of the study, it is crucial to undertake pre-processing procedures for the original dataset of APC and AVL measurements. When dealing with such raw data, these procedures are necessary to address various challenges that may arise and to establish the collected data's usability. In this research, we conducted a preliminary evaluation of data quality using a formal analysis procedure proposed in the literature. Subsequently, we refined the original raw dataset to encompass only high-quality data for further analysis. Additionally, data were augmented by integrating external data sources and manipulated to better fit the framework of the research. Subsection \ref{subsec:pre-processing} provides detailed methodologies and results of this pipeline.  

Following data cleaning, the research shifted its focus to the realm of prediction, which is then described next in Subsection \ref{subsec:forecasting}. At this stage, the primary objective was the construction and analysis of models for the identification of undercrowding with the application of statistical and machine learning tools.

\subsection{From raw APC data to analysis-ready input data}\label{subsec:pre-processing} 

\subsubsection{Data cleaning and pre-processing} \label{subsub:cleaning}

The initial phase of the pre-processing aimed to reconstruct the overall count of passengers boarding and alighting at each stop, along with the number of passengers on board between stops. It is noteworthy that the raw data were received in the form of signals from the system, with an uneven frequency, resulting in a varying number of signals available for each stop. The first step involved the creation of a new dataset structured with one row for each stop using variables related to the total count of passengers, boarding passengers, and alighting passengers contained in these signals. This revised dataset was designed to be more easily interpreted by public transport planners and served as input data for our model.

Before conducting any analysis with APC data, it was essential to address several challenges to ensure the usability and reliability of the collected data. 
In this study, we adopted the framework proposed by \citet{barabino2014offline}, which comprises three key phases: i) Matching and Parsing, ii) Validation, and iii) Anomaly Detection.

The Validation phase was crucial to ensure high-quality analysis of the data. It involved handling various aspects, including:
\begin{itemize}
    \item Noise: While raw data may offer correct information on people counting, the geo-localized information could be subject to inaccuracies;
    \item Missing data: Instances where no signals were registered for one or more stops along the bus route were addressed;
    \item Outliers detection: To maintain data integrity, we identified and dealt with raw data points that fell outside significant ranges.
\end{itemize}

Noise detection was conducted using only the data from APC. In the dataset, the "Noise" variable was used to identify noisy data. Missing data were identified by comparing the data from APC at the single ride level with the data from the timetabling schedule. The outliers detection phase relied on the people counting information within the APC data.

The registered noise was found to affect approximately $2.29$\% of the rides included in the sample. Analyzing the magnitude of noise in terms of the percentage of signals per ride that are affected, we observed that the average value per route is lower than $5$\% in $97.7$\% of the cases, and is on average equal to $0.89$\%.

For missing data, approximately $3.86$\% of the rides were affected. Among them, the value of missing values ranges between $10$\% and $95$\% with an average of $33$\%. 

The identification of outliers in the passenger counting data involved a thorough examination of the count data per vehicle. This process aimed to identify potential malfunctions in the APC systems installed on the vehicles. During this process, two types of anomalies were identified:
\begin{itemize}
    \item absence of both boarding and alighting passengers at each ride and every stop;
    \item absence of alighting passengers and presence of boarding passengers at each ride and every stop, and vice versa;
\end{itemize} 
Upon conducting the analysis, it was found that outliers in the APC data affected an average of $9$\% of vehicles per day. However, it is essential to note that $76.54$\% of the identified problematic vehicles were equipped with old-generation sensors that exhibited lower reliability compared to other sensors. 

In addition to analyzing outliers at the vehicle level, it was crucial to screen the data at the ride or even stop level to limit measurement errors, as emphasized by \citet{barabino2014offline}. To identify outliers in the boarding and alighting passenger data at the single-stop level, we employed a bag-plot methodology \citep{rousseeuw1999bagplot}. The process involved extracting a bivariate sample, consisting of the number of people boarded and the number of people alighted, at each stop. If the sample does not present any anomalous measurement (both people boarded and alighted are never more than a certain threshold, fixed at $50$), all the observations are classified as non-outliers. Otherwise, all the observations in the sample are classified according to a bagplot.

Following the application of the bagplot methodology, we incorporated additional domain knowledge provided by public transport operators to identify potential outliers that may not have been detected in the initial analysis. The results of this combined approach showed that outliers affected only $0.21$\% of the analyzed rides. Moreover, on average, these outliers impacted less than $10$\% of the stops per ride. 
 
The anomaly detection phase encompassed the identification of various anomalies, including departure/arrival from/at the depot, unexpected breakdowns, interrupted trips, and unplanned detours.
For departure/arrival from/at the depot, unexpected breakdowns, and interrupted trips, we utilized the service status information already included in the AVL data. This information enabled us to detect and classify these anomalies without the need for any specific pre-processing. The detection of unplanned detours required the integration of APC-AVL data with timetabling information to identify such cases.

Corrective actions were adopted to partially correct those data. Specifically, incomplete rides with more than $10$\% of missing values and rides with at least one noisy observation or at least one outlier observation were removed from the dataset. Moreover, when anomalies were detected, the full rides were excluded from the analysis.

\begin{figure}[t]
\centering
    \subfigure[Number of rides between weeks]{\resizebox*{6cm}{!}{\includegraphics{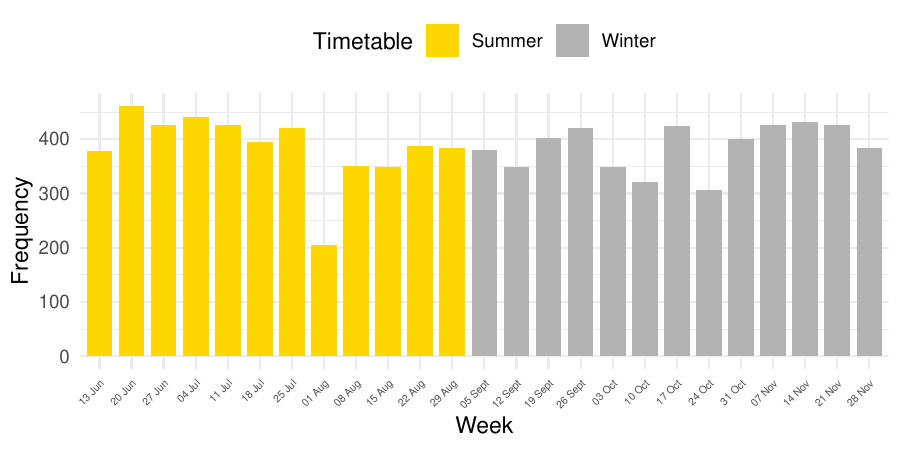}}}\hspace{0.25cm}
    \subfigure[Average number of rides per day type]{\resizebox*{6cm}{!}{\includegraphics{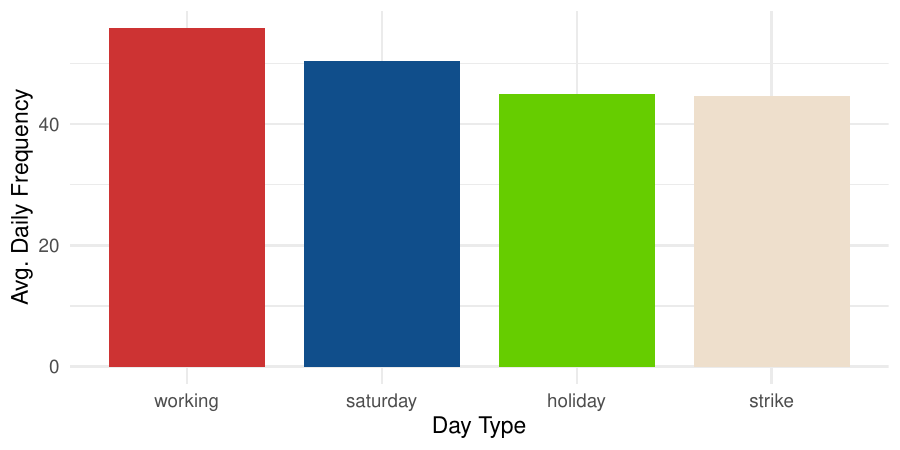}}}
    \caption{Distribution of the number of rides between weeks (a) and day types (b).}
    \label{fig:day}
\end{figure}

\begin{figure}[t]
  \centering
 \includegraphics[width=12cm]{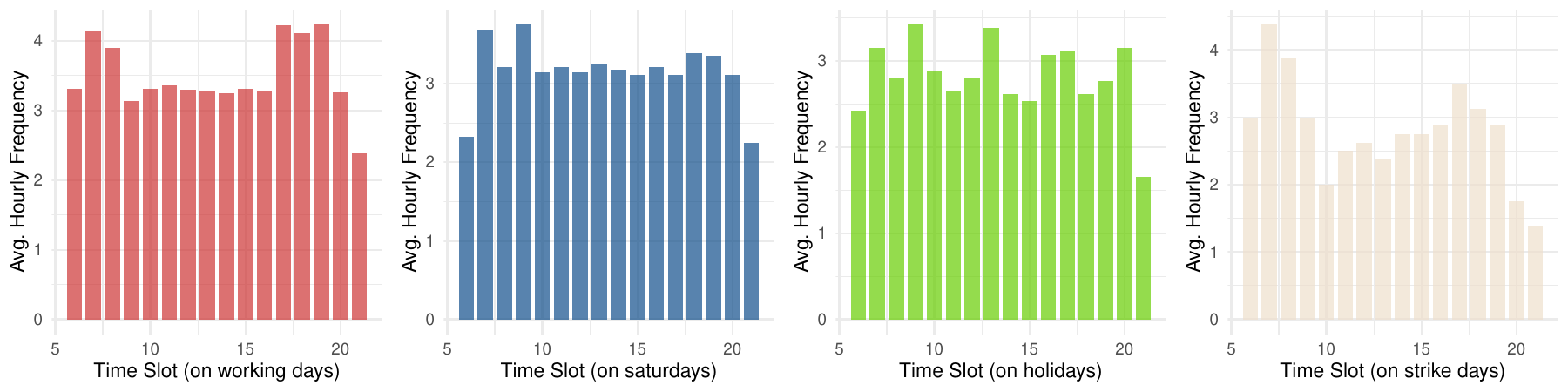}
  
  \caption{Average number of rides per time slot in different day types}
  \label{fig:hour}
\end{figure}

The final dataset consisted of $10,078$ rides traveling from the center to the suburbs. The distribution of the number of rides between weeks is shown in Figure \ref{fig:day} (a). The average number of rides per day type and time slot in different day types are illustrated in Figure \ref{fig:day} (b) and Figure \ref{fig:hour}. Compared to the original dataset of raw APC measurements, the elaborated version ensures higher reliability of the data and consequently allows for a more precise analysis of the urban mobility pattern and application to the study of critical situations.

\subsubsection{Data manipulation and integration}\label{subsubsec:augmentation}

The resulting dataset was further elaborated through the aggregation of the rides by hourly time slot. This step was essential to address the variability in service within the time slots, stemming from diverse programmed frequencies and unscheduled changes (e.g., vehicles with delays or ahead of schedule). This variability has the potential to impact the load factor \citep{van2015data}, necessitating this corrective measure. The aggregation of the rides by time slot has an immediate interpretation: all the rides in the same time slot were replaced with a single virtual aggregated ride that collects all the passengers transported in the hour and offers the same cumulative hourly capacity. The aggregated rides inherited the timing features (i.e., time slot and date) and the spatial features (i.e, the reference stop) from the rides of the original dataset, while aggregating the information of the vehicle capacity and the passenger counts of boarding, alighting, and on-board people.

Formally, let $D$ and $T$ be the set of dates and the interval of times: $D = \{06/06/2022, \dots, 05/12/2022\}$ and $T = [00\text{:}00\text{ am}, 11\text{:}59\text{ pm}]$. Let also $S$ be the set of hourly time slots defined as closed sub-intervals in the set of times, $S=\{[0\text{:}00\text{ am}, 0\text{:}59 \text{ am}],\dots, [11\text{:}00\text{ pm},11\text{:}59 \text{ pm}]\}$. Given a ride that runs on day $d \in D$ at time $t \in T$, there can be defined $C_{t,d}$ as the capacity of the ride, and $O_{t,d}^j$ as the occupancy of the ride in the $j-$th segment of the route. Then, the aggregate ride $i$ is defined as that with a capacity $C_i$ equal to the sum of the capacities $C_{t,d}$ of the rides in a specific time slot $s^\star\in S$ and day $d^\star \in D$:

\begin{equation}\label{eq:C}
    C_i = C_{\{s^\star, d^\star\}_i} = \sum_{t\in s^\star, d=d^\star} C_{t,d} 
\end{equation}

and with occupancy per segment $j$ equal to the sum of occupations per segment $O_{t,d}^j$ of the rides in that time slot $s^\star\in S$ and day $d^\star \in D$, that is:
\begin{equation}\label{eq:O}
   O_{i}^j = O_{\{s^\star, d^\star\}_i}^j = \sum_{ t \in s^\star, d=d^\star } O_{t,d}^j 
\end{equation}

It should be noted that the capacity of the aggregate ride is constant along the ride, and therefore does not depend on the segment $j$, unlike the occupancy that varies on each segment between stops due to the presence of passengers boarding and alighting. 

Aggregate rides were generated for each combination of time slot and date $\{s,d\}_i \subset S\times D $ present in the analyzed dataset, that is, removing all the combinations where no rides were running. This phase led to the selection of $50,076$ observations belonging to $N_{r}\!=\!2,782$ aggregate rides on the route, each ride measured at $N_s\!=\!18$ segments between stops.

The definition of Load Factor, measured at each segment, was extended for an aggregate ride $i$ and a segment $j$ as follows:

\begin{equation}\label{eq:LF}
   LF_{i}^j = \frac{O_{i}^j}{C_i}
\end{equation}

The Load Factor of aggregate rides has been used as an indicator of undercrowding or overcrowding situations, serving either as a measure of efficiency or effectiveness. Load Factor values below a certain threshold $c_{low}$ correspond to situations of low efficiency because of undercrowding and, vice versa, Load Factor values above a certain threshold $c_{high}$ are indicators of discomfort resulting from overcrowding. Given the characteristics of the examined route, the undercrowding situation was investigated. The latter is defined as a Bernoulli random variable $y_{ij}$ whose value, for each aggregate ride $i$ between the $j$-th and $(j\!+\!1)$-th stops, as in Equation \ref{eq:Y}, depends on the value of Load Factor and the undercrowding threshold chosen:
\begin{equation}\label{eq:Y}
    y_{ij} = \begin{cases} \: 1 \text{       if } LF_{i}^j \leq c_{low}\\ \: 0 \text{       otherwise.} \end{cases}
\end{equation}

The dataset was then augmented by incorporating additional data pertaining to meteorological conditions, sourced from the Historical Weather archive of open-meteo.com \citep{zippenfenig2023open-meteo}. These data encompassed various parameters including temperature, wind speed, cloud coverage, relative humidity, and precipitations. Moreover, information on the scheduled service was retrieved from the transportation company. Each day was classified according to the rendered programmed schedule, taking into account the type of service day and the seasonal service. This categorization included different scenarios: regular workdays that feature the standard baseline service, workdays influenced by labor strikes that lead to non-guaranteed service, Saturdays and holidays with a reduction in service.
An additional classification consisted in splitting the period of interest in the Winter season and the Summer season, among which the service varies. However, this service type variable was not further considered since its information is already inherently included in the Week variable.
All the available data that were of interest for subsequent analyses are summarized in Table \ref{tab:input_data}.

\begin{table}[htb]
    \centering
    \scriptsize
    \caption{Input data description}
    \begin{tabular}{ccp{3.8cm}p{4.2cm}}

        \toprule
        \textbf{Data} & \textbf{Variable} & \textbf{Variable description} & \textbf{Summary statistics} \\
        
        \midrule

        APC & Undercrowding& Binary variable $y_{ij}$ with value 1 for under-crowded situations, 0 otherwise& 0 [85.7\%], 1 [14.3\%]\\ 
        & Time slot & Hour of the aggregate ride & [6:00 AM -- 9:00 PM]\\
        & Date & Date of the aggregate ride, from which to extract the information on the day number, week number& [2022-06-06 -- 2022-12-04]\\
        & Segment & Segment between stops of the line & [1 -- 18]\\
    
        \midrule
        
        Weather & Temperature &  Average temperature [°C] & [0.7--36.6] mean=20.9 sd=7.79 \\
        & Wind speed & Average velocity of the wind [km/h] & [0.4--20.9] mean=6.19 sd=3.43\\
         & Cloud coverage & 
        Percentage of cloud coverage [\%] & [0--100] mean=33.6 sd=32.52\\
        & Humidity &  Relative humidity [\%] & [21.0--100]  mean=63.9 sd=20.02 \\
        & Rain & Average amount of precipitation [mm] & [0--8.2] mean=0.01, sd=0.4971 \\
        
        \midrule

        Service & Day type & Type of service day & Working days [65.8\%], working days with strikes [4.2\%], Saturdays [14.9\%], holidays [15.1\%] \\
        & Service season & Type of service season & Summer [49.9\%], winter [50.1\%] \\ 
        \bottomrule

    \end{tabular}
    \label{tab:input_data}
\end{table}

\subsection{Prediction models}\label{subsec:forecasting}

The subsequent step aimed to estimate the probability of observing undercrowding from the input data consisting of the cleaned and augmented APC dataset described in Subsection \ref{subsec:pre-processing}. The analysis of undercrowding involved the construction and training of statistical and machine-learning methods, aimed at estimating the probability of undercrowding based on temporal factors, service parameters, and meteorological conditions. Two models were under consideration: the statistical approach known as the Generalized Linear Mixed-Effect Model \citep{nelder1972generalized} and the algorithm of Generalized Mixed-Effect Random Forest \citep{Pelegatti2021Generalized}. A comparative analysis was conducted, evaluating the models' performance in terms of prediction accuracy and interpretability of the results. 
Then, the predictions of the chosen model were examined to gain insight into the pattern of undercrowding for the route under analysis.

\subsubsection{Generalized Linear Mixed-effects Model} \label{subsubsec:GLMM}
Within the statistical setting, a Generalized Linear Model (GLM) could be considered to estimate the probability of encountering undercrowding (response) with respect to the other variables for each aggregate ride $i$ in each segment $j$ of the route \citep{nelder1972generalized}.
In this research, the generalization of simple linear models through GLMs is suitable due to the distribution of the response $y_{ij}$, which assumes values in $\{0,1\}$ and can be modeled as a random variable following a Bernoulli distribution related with the covariates through the logit link function. 


Other studies interested in the same response variable chose to summarise it with a single scalar value per ride. For example, one can choose the response to be $\prod_j y_{ij}$ to single out the presence of undercrowding on all segments of the route, and $\sum_j y_{ij}$ to count the number of segments suffering undercrowding in a route. In \citet{barabino2014offline}, the authors chose to study only the most representative segment of the route, while other researchers such as \citet{hu2018bus} studied the cumulative passenger load over the whole route. These approaches may be able to quantify seasonality effects on the demand but lose the information on the distribution of undercrowding situations along the route, i.e. the identification of the most critical segments between stops. 

In the proposed methodology the spatial and temporal granularity of the data was kept, considering the full set of $\{y_{ij}\}_{ij}$ and including the “segment” factor. Given that the segment factor is linked to a categorical variable with numerous levels, it is more advantageous to treat it as a random effect rather than a fixed one, as commonly observed in literature. Hence, a Generalized Linear Mixed-effects Model (GLMM) was preferred and used \citep{venables2002lmm}. The GLMM included both fixed and random effects in the linear predictor, moving forward from the GLM with only fixed-effects. The response variable was hence estimated as a function of the linear combination of a random-effect term $z_{0j}$, added to the model to account for the variation between segments, and fixed-effects of the remaining covariates.

The regressors considered in the model were: the type of service day, included as dummy variables for working days, strikes, Saturdays, and holidays; the time slot of the ride, as polynomial terms of degree $D_s$; the week number, as polynomial terms of degree $D_w$; interaction terms between them; the weather data of temperature, precipitation, cloud coverage, wind velocity, and relative humidity; a random intercept different for each segment ($z_{0j}$).
The choice of using polynomials for both time slot and week number started from the preliminary descriptive analysis, where a nonlinear relationship between the Load Factor and these variables has been outlined. The need for a more complex parametrization has then been supported by the comparison of the model performances for the two cases of simple linear and polynomial linear models. The dataset was randomly divided into a training set and a test set, comprising 70\% and 30\% of the rides, respectively. The division is performed based on rides rather than individual observations, aligning with the concept that in a real-world scenario, when a new observation emerges, the measurements for all other data points within the same ride become accessible. Among different degrees for the time slot variable, $D_s\in\{0, \dots, 10\}$, the optimal one was found by minimizing the Mean Squared Error (MSE) of the 10-fold Cross-Validation (CV) on the training set when fixing the degree of the week variable at a predefined value, $D_w=3$. Subsequently, after fixing $D_s$ equal to its optimal value previously found, different degrees $D_w\in\{0,\dots,6\}$ were evaluated and the one minimizing the 10-Fold CV MSE on the training set was chosen.
The model with the lowest value of MSE consisted in the case of the polynomial linear model, with the two polynomial terms of time slot and week number of the $D_s$-th and $D_w$-th degree. The formulation of the proposed logistic regression model with mixed-effect for the Bernoulli random variables $y_{ij}$ is expressed in the following:


\begin{equation}
    \begin{split}
    y_{ij} \sim & \: Bernoulli (\eta_{ij})\\
    logit(\eta_{ij}) = & 
        \: \beta_0 
        + \sum_{k=1}^{D_s} \beta_{1d}\, x^k_{1i} 
        + \sum_{k=1}^{D_w} \beta_{2d}\: x^k_{2i} 
        + \sum_{p=3}^{P} \beta_p\, x_{pi} 
        + z_{0j} \\    
    z_{0j} \sim & \: \mathcal{N}(0, \sigma_{z}^2)\\
    \end{split}
    \label{eq:mod_glm}
\end{equation}

\noindent where $y_{ij}$ denotes the undercrowding of the aggregate ride $i\in\{1,\dots,N_r\}$ in the segment $j\in\{1,\dots,N_r\}$; $P$ is the number of regressors considered; the $(P+D_s+D_w-1)$-dimensional vector of parameters includes the intercept $\beta_0$, the parameters of time slot $\bm{\beta}_1=[\beta_{1,1},\dots,\beta_{1,D_s}]$, the parameters of the week $\bm{\beta}_2=[\beta_{2,1},\dots,\beta_{2,D_w}]$, and the parameters of other regressors $\bm{\beta}=[\beta_{3},\dots,\beta_{P}]$; $x_{ip}$ represents the value of the $p$-th variable, in particular $x_{1}$ is the variable of time slot up to degree $D_s$, and $x_{2}$ is the variable of week number up to degree $D_w$; $z_{0j}$ is the random intercept for segment $j$.
The model trained on the training set was then evaluated on the test set, as previously defined.

\subsubsection{Generalized Mixed-Effects Random Forest}\label{subsubsec:GMERF}

As an alternative to the GLMM approach, an ensemble-based method as a Random Forest (RF) was trained to classify situations of undercrowding based on the available covariates \citep{breiman2001random}. RFs have become popular because of their ability to outperform statistical linear models for their ability to capture possibly non–linear dependencies between input and output variables. Moreover, they implicitly perform feature selection and are less prone to overfitting. Although RFs have always been questioned for their interpretability because of their complex decision tree ensemble, recent techniques of interpretable machine learning can provide insights into the inner workings of RFs, making them more interpretable and transparent.

Following the same intuitions that lead to the GLMM, instead of a simple RF, this work made use of a Generalized Mixed-Effect Random Forest (GMERF). Proposed by \citet{Pelegatti2021Generalized}, GMERF extends traditional random forests to handle mixed-effects models in the case of binary output. It incorporates both a fixed effect and a random effect part into the model structure, making it suitable for analyzing hierarchical data. This makes GMERF suitable for modeling complex relationships in data, as an RF would do, as well as considering individual and group-level effects of grouped observations. The GMERF model formulation applied in the analysis takes the following form:

\begin{equation}
    \begin{split}
    y_{ij} \sim & \: Bernoulli (\eta_{ij})\\
    logit(\eta_{ij}) = & \: f(\bm{x}_i) + z_{0j}\\    
    z_{0j} \sim & \: \mathcal{N}(0, \sigma_{z}^2)\\
    \end{split}
    \label{eq:mod_gmerf}
\end{equation}

Here, the response variable $y_{ij}$ represents the indicator of undercrowding. Being a binary variable, it is again modeled as a Bernoulli random variable. The fixed effects component denoted as $f(\bm{x}_{i})$, is no longer presumed to have a linear nature. 
All the available variables described in Table \ref{tab:input_data} were included as fixed predictors in the RF model: time slot, week, day type, service season, and all weather factors. Simultaneously, the random part consisted of a random intercept $z_{0j}$ that considered the affiliation with particular segments of the line.
The model was initially trained and then tested on the same subsamples used for the GLMM. As initial parameters, random effects are set to be equal to the estimated ones in the previous GLMM.

\section{RESULTS}  \label{sec:results}

The threshold for defining undercrowding was established considering the characteristics of the examined route. A Load Factor of $t_{low}\!=\!1$\% was suggested by the public transport operator as the threshold for identifying undercrowding on this specific route. It's noteworthy that the choice of this threshold can be subject to modification based on specific needs or circumstances, allowing for adaptability to changing conditions or requirements.
Given this threshold, the two models were run to estimate the probability of having situations of undercrowding for week, service type of day, time slot, segment of the route, and weather data. 

\begin{figure}[tb]
\centering
    \subfigure[Different values of $D_s$ and fixed $D_w=3$.]{\resizebox*{6cm}{!}{\includegraphics{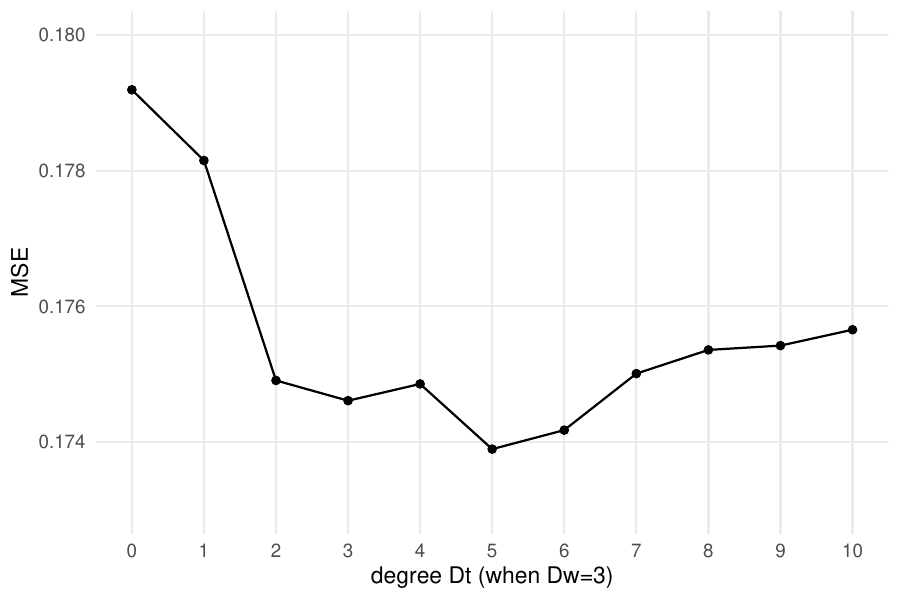}}}\hspace{0.5cm}
    \subfigure[Different values of $D_w$ and fixed $D_s=5$.]{\resizebox*{6cm}{!}{\includegraphics{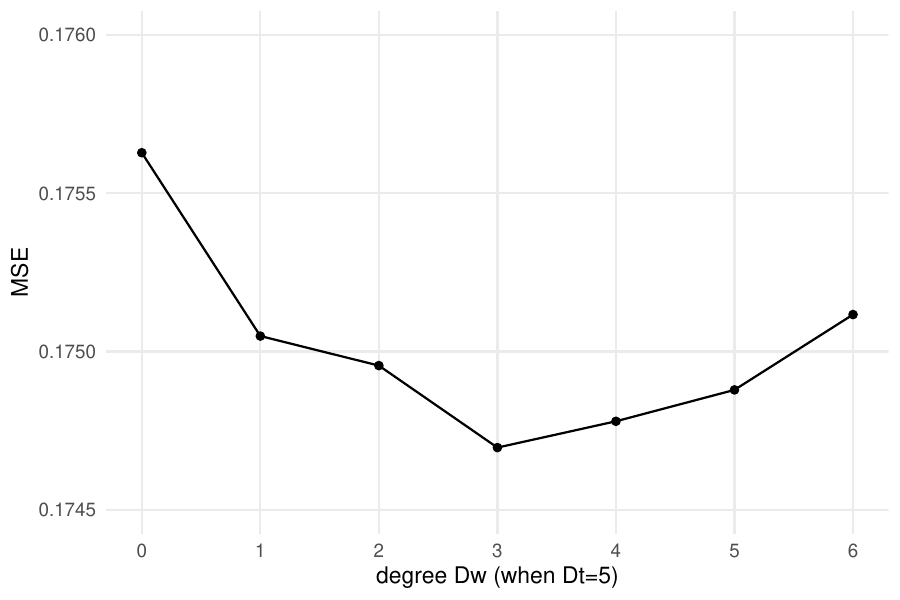}}}
    \caption{10-Fold CV MSE for different degrees of time slot $D_s$ (a) and for different degrees of week $D_w$ (b) in Model \ref{eq:mod_glm}.}
    \label{fig:MSE_tw}
\end{figure}

For what concerns the GLMM, in the initial phase, the optimal hyperparameters $D_s$ and $D_w$ were determined by minimizing the 10-Fold CV MSE on the training dataset in a two-step process. Firstly, maintaining the week variable's degree at a predefined value of $D_w = 3$, the analysis revealed that $D_s = 5$ minimized the 10-Fold CV MSE, as illustrated in Figure \ref{fig:MSE_tw} (a). Subsequently, with $D_s$ fixed at the optimal value of 5, found in the previous step, the optimal degree for the week variable, minimizing the MSE, was found to be $D_w = 3$, as shown in Figure \ref{fig:MSE_tw} (b). Secondly, feature selection was performed to remove uninformative covariates. Both the temperature, the cloud coverage and the relative humidity were tested to be not significant for the model (for CloudCoverage, $pval\!=\!0.73$; for RelativeHumidity, $pval\!=\!0.94$) and were hence removed in the final model formulation.

\begin{figure}[tb]
    \centering
    \includegraphics[width=6cm]{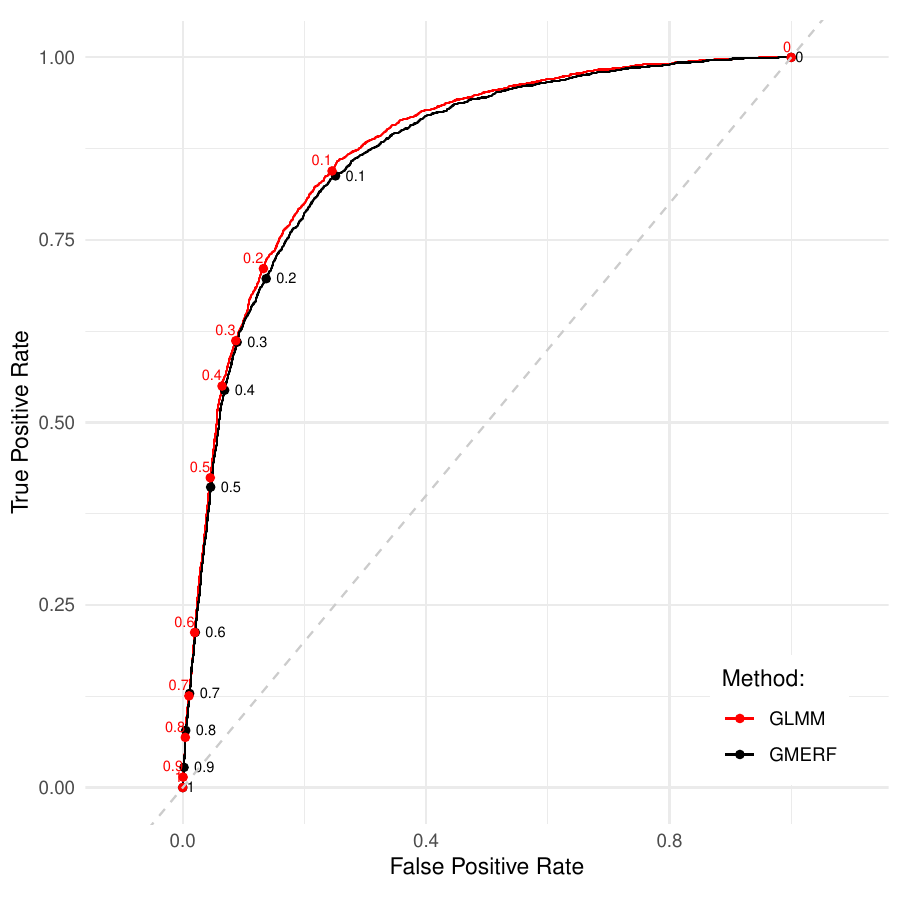}
    \caption{ROC curve evaluated on the test set for both Model \ref{eq:mod_glm} (in red) and Model \ref{eq:mod_gmerf} (in black). }
    \label{fig:ROC}
\end{figure}

Both the GLMM and the GMERF models were validated on a test set of $30$\% of the rides randomly selected. The two ROC curves depicted in Figure \ref{fig:ROC} refer to the output of the GLMM (in black) and the GMERF (in red). They show that both methods provide a good balance of true positive rate (TPR) and false positive rate (FPR) for different classification thresholds, with Area Under the Curve $AUC_1\!=\!0.868$ for the GLMM approach and $AUC_2=0.8676$ in the GMERF approach. In the former, the classification threshold of $f\!=\!0.5$ leads to an accuracy of $acc_1=87.8$\% in identifying undercrowding, and similar results in accuracy are obtained from the GMERF, with an accuracy of $acc_2=87.7$\%. Nevertheless, the methods slightly differ in the way they misclassify observations: as shown in the misclassification Table \ref{tab:misc_GLMM}, GLMM tends to misclassify less critical situations in non-undercwording when compared to the GMERF. 

Regarding the predictors which significantly influence the classification, temporal factors, their interactions, and the three weather covariates result marginally significant for the GLMM ($pval\!<\!0.01$). Furthermore, the model is characterized by a high Proportion of Variability of the Random Effects (PVRE), which reaches $PVRE_1=35.9$\% underlying the importance of including the 'segment' grouping factor in the model from a spatial perspective.
Instead, for the GLMERF, the variability explained by random effects is limited to $PVRE_{2}=28.6$\% that, although smaller than in the GLMM case, still shows the relevance of adding the segment factor.

\begin{table}[htb]
\centering
    \caption{Misclassification table of Model \ref{eq:mod_glm} (left) and Model \ref{eq:mod_gmerf} (right) measured on the test set for classification threshold $f\!=\!0.5$}
    \label{tab:misc_GLMM}%
    
    \begin{tabular}{l|cc|cc|}
    \hfill Pred& \multicolumn{2}{|c|}{GLMM}& \multicolumn{2}{|c|}{GMERF}  \\
    True& $\hat{y}=1$&$\hat{y}=0$& $\hat{y}=1$&$\hat{y}=0$ \\
        \hline
    $y=1$& $862$ ($5.7$\%)& $594$ ($4.0$\%)& $690$ ($4.6$\%)& $1\,405$ ($9.4$\%)\\
    $y=0$& $1\,233$ ($8.2$\%)& $12\,341$ ($82.1$\%)& $441$ ($2.9$\%)& $12\,494$ ($83.1$\%)\\ 
    \hline

    \end{tabular}

\end{table}

By looking at these performance measures, it emerges that the two models have comparable ability for the purpose of the presented application case. They discern undercrowding situations with similar accuracy and exploit the information of the same covariates for predicting the outcome of new observed situations. Among the two, GLMM has been chosen to present the further analysis of the present research because of the similar results and the easier interpretability of the model outcome. In the following, the results of the GLMM model trained on the full sample of observations are presented and commented. 

\begin{figure}[h]
      \centering
     \includegraphics[width=7cm]{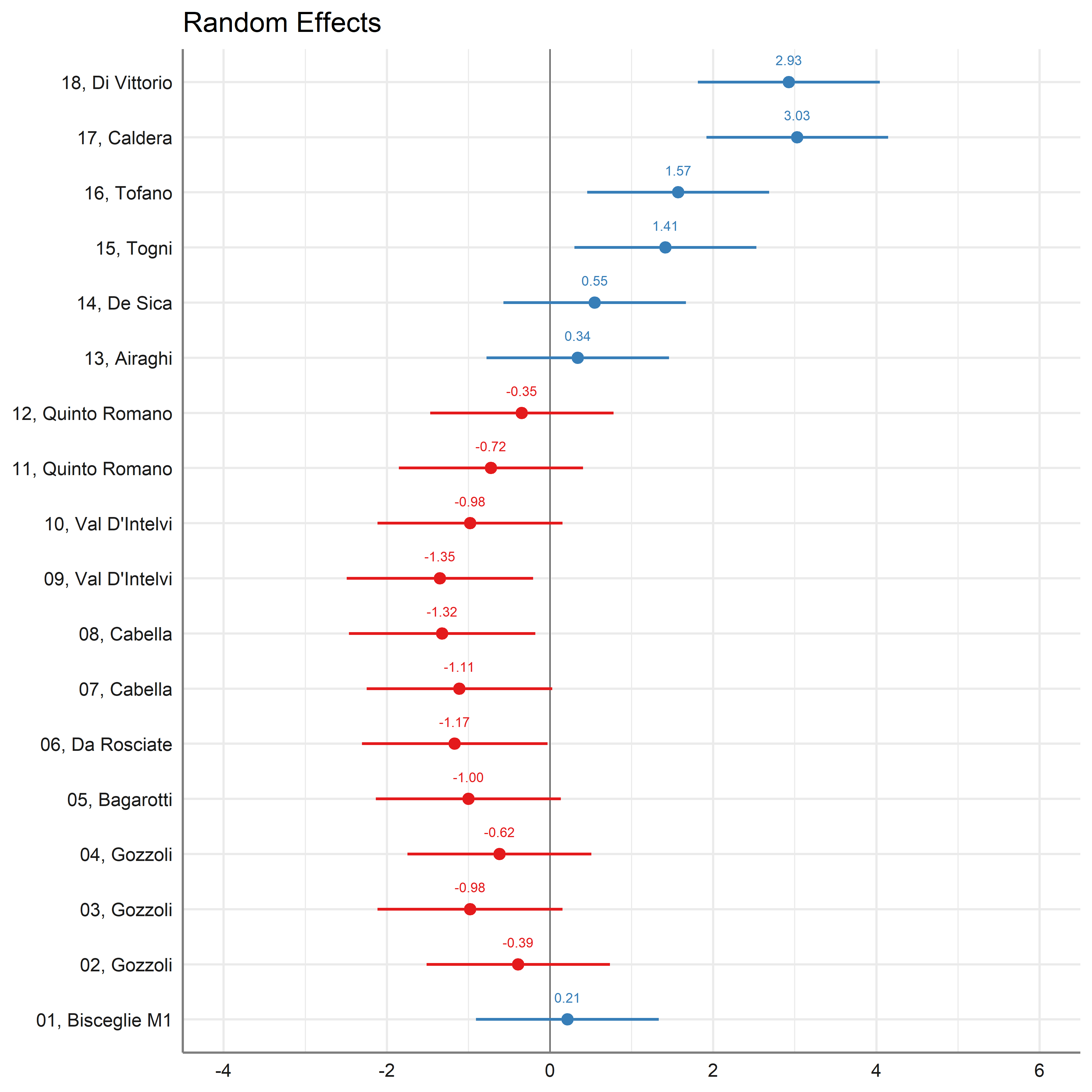}
      
      \caption{Random effects}
      \label{fig:random_effects}
   \end{figure}

\begin{table}[htb]
    \centering
    \resizebox{\columnwidth}{!}{
    \begin{tabular}{lcccccc}
    \toprule
     & Estimate & Estimate\_CI\_0.05 & Estimate\_CI\_0.95 & SE & z & p \\ 

    \midrule
    (Intercept) & -2.79 & -3.47 & -2.10 & 0.35 & -7.97 & 0.00 \\ 

  Day type [Saturdays] & 0.44 & 0.35 & 0.53 & 0.04 & 9.81 & 0.00 \\ 
  Day type [Holidays] & 1.15 & 1.07 & 1.23 & 0.04 & 28.04 & 0.00 \\ 
  Day type [Strikedays] & 0.64 & 0.47 & 0.81 & 0.08 & 7.55 & 0.00 \\ 
  
  Time slot [\^{}1] & 64.66 & 54.14 & 75.19 & 5.37 & 12.04 & 0.00 \\ 
  Time slot [\^{}2] & 58.23 & 48.28 & 68.18 & 5.08 & 11.47 & 0.00 \\ 
  Time slot [\^{}3] & 48.78 & 39.89 & 57.66 & 4.53 & 10.76 & 0.00 \\ 
  Time slot [\^{}4] & 44.80 & 35.97 & 53.63 & 4.50 & 9.95 & 0.00 \\ 
  Time slot [\^{}5] & 16.21 & 7.44 & 24.97 & 4.47 & 3.62 & 0.00 \\
  
  Week [\^{}1] & 19.74 & -0.09 & 39.58 & 10.12 & 1.95 & 0.05 \\ 
  Week [\^{}2] & -18.48 & -29.69 & -7.27 & 5.72 & -3.23 & 0.00 \\ 
  Week [\^{}3] & 23.25 & 14.00 & 32.51 & 4.72 & 4.92 & 0.00 \\ 
  
  Precipitation & 0.17 & 0.11 & 0.22 & 0.03 & 5.89 & 0.00 \\ 
  WindSpeed & -0.02 & -0.03 & -0.01 & 0.01 & -4.05 & 0.00 \\ 
  
  Time slot [\^{}1] * Day type [Saturdays] & -63.19 & -81.62 & -44.75 & 9.40 & -6.72 & 0.00 \\ 
  Time slot [\^{}2] * Day type [Saturdays] & 12.68 & -5.85 & 31.22 & 9.46 & 1.34 & 0.18 \\ 
  Time slot [\^{}3] * Day type [Saturdays] & -33.07 & -51.80 & -14.33 & 9.56 & -3.46 & 0.00 \\ 
  Time slot [\^{}4] * Day type [Saturdays] & -35.59 & -54.33 & -16.84 & 9.56 & -3.72 & 0.00 \\ 
  Time slot [\^{}5] * Day type [Saturdays] 
  & -20.58 & -39.43 & -1.73 & 9.62 & -2.14 & 0.03 \\ 
  Time slot [\^{}1] * Day type [Holidays] & -139.53 & -155.97 & -123.09 & 8.39 & -16.64 & 0.00 \\ 
  Time slot [\^{}2] * Day type [Holidays] & 93.53 & 76.53 & 110.52 & 8.67 & 10.79 & 0.00 \\ 
  Time slot [\^{}3] * Day type [Holidays] & -96.41 & -113.42 & -79.40 & 8.68 & -11.11 & 0.00 \\ 
  Time slot [\^{}4] * Day type [Holidays] & -43.38 & -60.25 & -26.50 & 8.61 & -5.04 & 0.00 \\ 
  Time slot [\^{}5] * Day type [Holidays] & -18.19 & -35.06 & -1.32 & 8.61 & -2.11 & 0.03 \\ 
  Time slot [\^{}1] * Day type [Strikedays] & -6.55 & -40.46 & 27.36 & 17.30 & -0.38 & 0.71 \\ 
  Time slot [\^{}2] * Day type [Strikedays] & -60.40 & -92.95 & -27.84 & 16.61 & -3.64 & 0.00 \\ 
  Time slot [\^{}3] * Day type [Strikedays] & 10.41 & -22.60 & 43.42 & 16.84 & 0.62 & 0.54 \\ 
  Time slot [\^{}4] * Day type [Strikedays] & 5.67 & -27.32 & 38.66 & 16.83 & 0.34 & 0.74 \\ 
  Time slot [\^{}5] * Day type [Strikedays] & -41.46 & -74.74 & -8.18 & 16.98 & -2.44 & 0.01 \\ 
  
  Day type [Saturdays] * Week [\^{}1] & -9.72 & -29.78 & 10.34 & 10.24 & -0.95 & 0.34 \\ 
  Day type [Holidays] * Week [\^{}1] & 18.45 & 0.52 & 36.38 & 9.15 & 2.02 & 0.04 \\ 
  Day type [Strikedays] * Week [\^{}1] & -27.16 & -57.17 & 2.85 & 15.31 & -1.77 & 0.08 \\ 
  Day type [Saturdays] * Week [\^{}2] & -13.54 & -33.44 & 6.36 & 10.15 & -1.33 & 0.18 \\ 
  Day type [Holidays] * Week [\^{}2] & -21.23 & -39.15 & -3.30 & 9.15 & -2.32 & 0.02 \\ 
  Day type [Strikedays] * Week [\^{}2] & -39.45 & -78.73 & -0.17 & 20.04 & -1.97 & 0.05 \\ 
  Day type [Saturdays] * Week [\^{}3] & 28.70 & 9.13 & 48.28 & 9.99 & 2.87 & 0.00 \\ 
  Day type [Holidays] * Week [\^{}3] & -25.05 & -42.85 & -7.26 & 9.08 & -2.76 & 0.01 \\ 
  Day type [Strikedays] * Week [\^{}3] & 13.51 & -20.38 & 47.41 & 17.29 & 0.78 & 0.43 \\ 
   \bottomrule
 
    \end{tabular}
    }
    \caption{Summary of the GLMM. For each fixed-effect term in the formulation of the model, there are reported the estimated value, the standard error, the test statistic z, and its p-value. The class of categorical variables is indicated between square parenthesis. The degree of the polynomial for Time slot and Week is reported between square parenthesis. }
    \label{tab:estim}
\end{table}

\begin{figure}[tb]
      \centering
     \includegraphics[width=12cm]{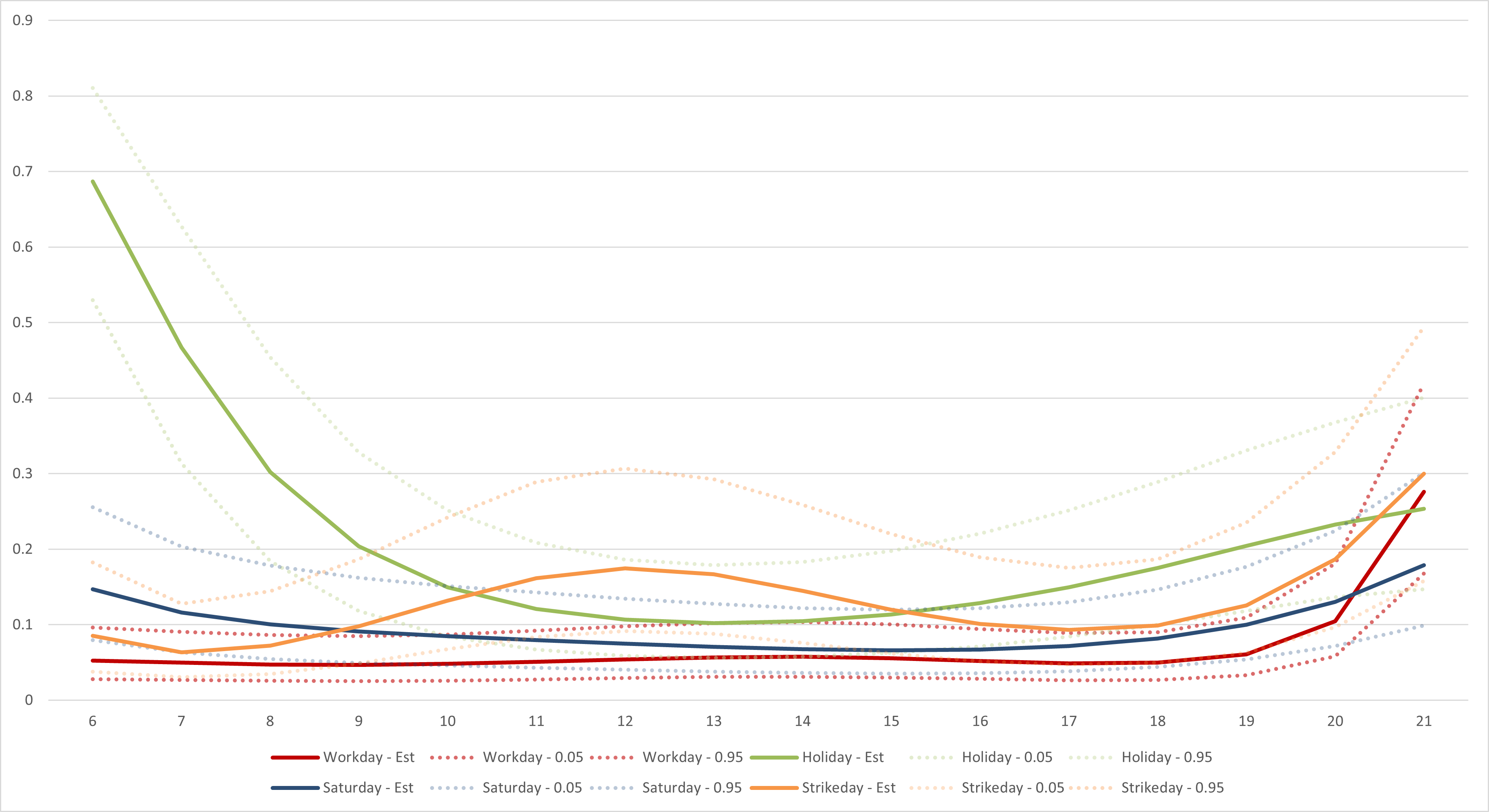}
      
      \caption{Marginal effect of Time slot and Day type on the probability of undercrowding. x axis: Time slot; y axis: probability of undercrowding; colors represent different Day types. While straight lines represent the pointwise estimation, the dashed lines represent the boundaries of the 95\% confidence intervals.} 
      \label{fig:marginal_effect_time}
\end{figure}

\begin{figure}[tb]
      \centering
     \includegraphics[width=12cm]{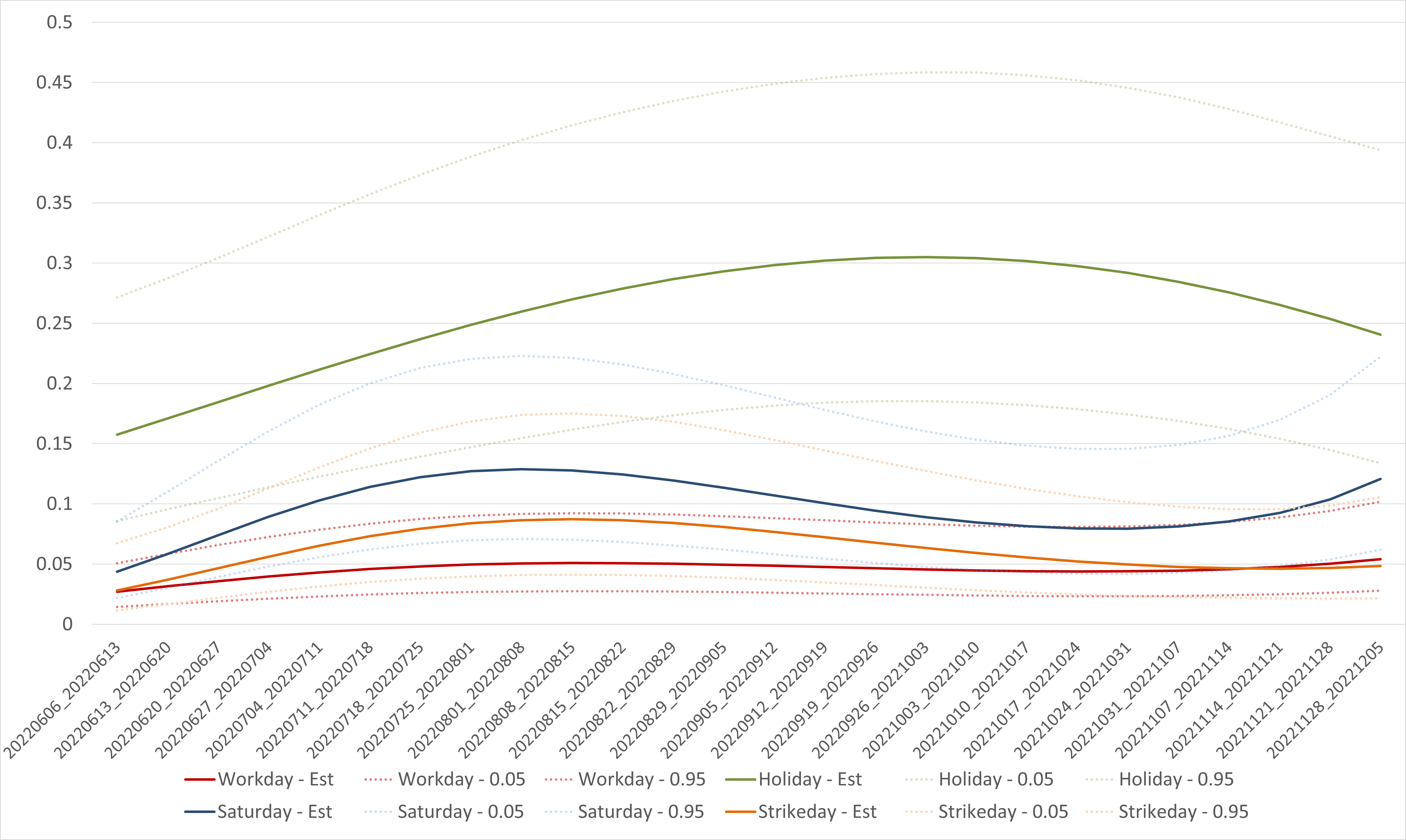}
      
      \caption{Marginal effect of Week and Day type on the probability of undercrowding. x axis: Week; y axis: probability of undercrowding; colors represent different Day types. }
      \label{fig:marginal_effect_week}
\end{figure}

The estimated random effects of the GLMM are depicted in Figure \ref{fig:random_effects}. These estimates outline that some segments have a larger probability to have undercrowding measurements, namely, the segment starting at the initial terminus and ones at the end of the route path (from station 13 to station 19).
Table \ref{tab:estim} shows the summary output of the model concerning the fixed-effects. 
Both weather covariates influence the probability of undercrowding, with odds ratios of $1.18$ for Precipitation and $-0.02$ for WindSpeed. This means that for an increase of 1 mm of Precipitation the probability of undercrowding increase of $+18$\%, while for an increase of 1 km/h of Windspeed the odds ratio decreases of $-2$\%.  
Regarding the polynomial terms of Time slot and Week, and their interaction with Day type, the table reports they are all significant. An easier explanation of their relationship with the probability of undercrowding is done by means of the plot of their marginal effects in Figure \ref{fig:marginal_effect_time} and Figure \ref{fig:marginal_effect_week}. 
From the former, it can be seen that undercrowding is more probable on Holidays than on other day types. Moreover, Holidays show a higher probability of undercrowding during the early morning hours (from 6am to 9am) and from the late afternoon (from 5pm to 10pm). On Strikedays the undercrowding is highly probable during the morning hours, which often correspond to the duration of the strike. Finally, during late hours of the day (from 8pm to 10pm) the probability of undercrowding slightly increases independently on the Day type. 
Regarding the relationship between Week and undercrowding in Figure \ref{fig:marginal_effect_week}, the trend of undercrowding probability is peculiar for Holidays, showing an increase for the winter months with respect to summer. Contrarily, the undercrowding in other Day types is estimated to be similar among the Weeks without an evident pattern. 

The model's results were then interpreted to identify undercrowding at the ride level. Integrating a segment-level analysis with a ride-level analysis provides a more comprehensive perspective, facilitating informed decision-making for the optimization and sustainable operation of public transportation systems. Different interventions can be indeed implemented to cope with critical situations localized in a specific segment, or spread throughout the journey, tailored to solve the specific problem. This approach not only facilitates the investigation of critical load factors at the entire ride level. It also enables the analysis of the impact of temporal factors on load factors at the ride level, controlling other external factors. Indeed, unlike a simple frequency counting method, the employed methodology allows for the examination of the spatio-temporal distribution of undercrowded rides while isolating the influence of weather factors and specific event data information. As a result, the model emerges as a versatile tool that extends beyond traditional descriptive analysis, opening avenues for scenario exploration by manipulating input data to assess the probability of undercrowding under varying conditions.  

\begin{figure}[tb]
      \centering
     \includegraphics[scale=0.5] {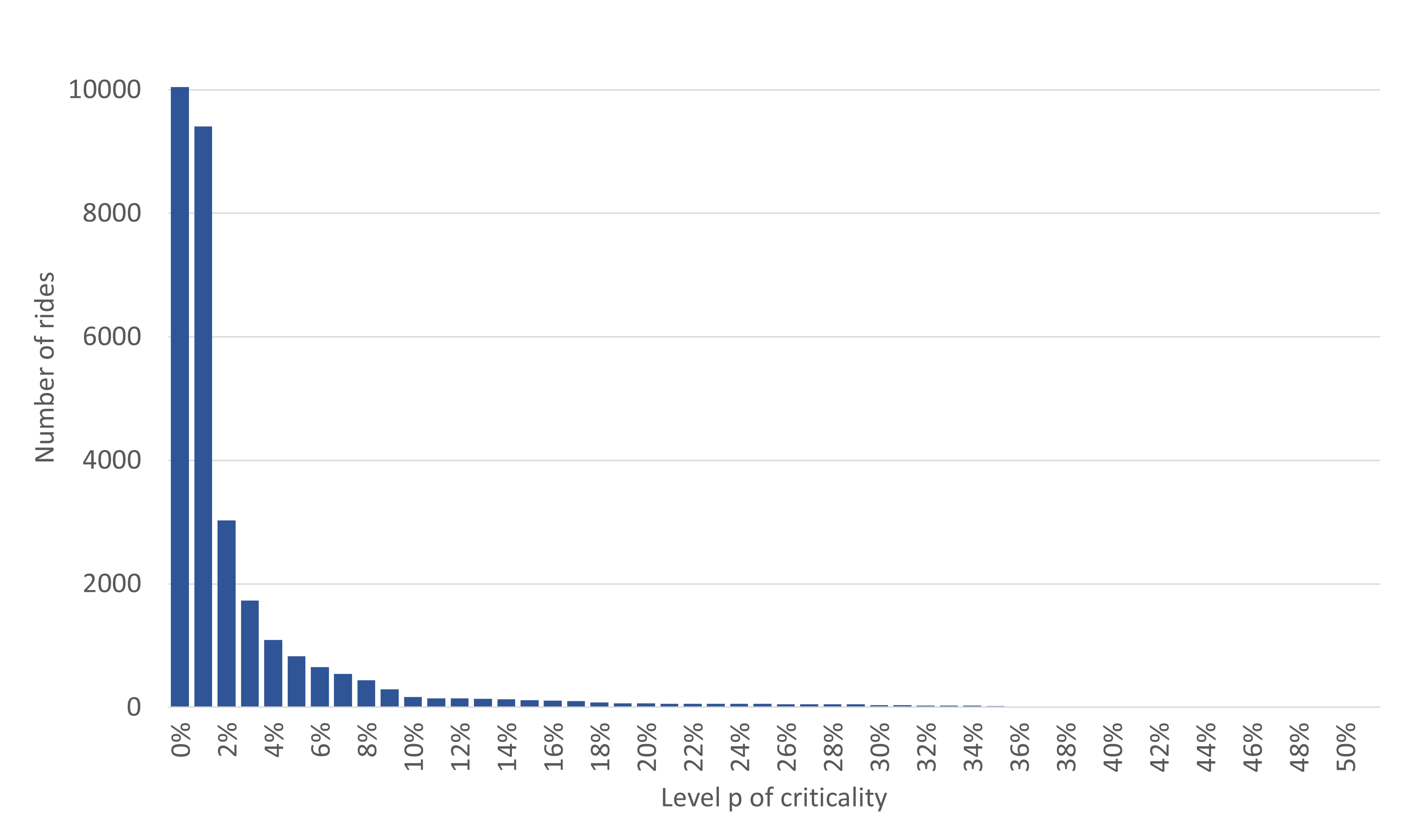}
      
      \caption{Number of fully undercrowded rides at level $p$, with $p$ varying in $[0,1]$}
      \label{fig:ride_p}
   \end{figure}

By establishing an upper-bound limit on the probability of encountering under-crowded scenarios, our ride-level analysis aids in evaluating the number of rides associated with a probability of undercrowing surpassing the set limit across the entire route. A ride is deemed undercrowded at level $p$, where $p$ falls within the range of $[0,1]$, if, in each of its segments, the probability of undercrowding is greater than or equal to $p$. The graph in Figure \ref{fig:ride_p} illustrates, for values of $p$ in the interval $[0,1]$, the count of fully undercrowded rides (y-axis) at level $p$ (x-axis) among the $10,046$ rides considered in the analysis outlined in this paper. The number of undercrowded rides rapidely decays until p=0.05, reaching $n_{0.05}=826$ rides, then slowly decreases until reaching almost zero at p=0.1 ($n_{0.1}=169$).

\begin{figure}[tb]
      \centering
      \includegraphics[scale=0.6]{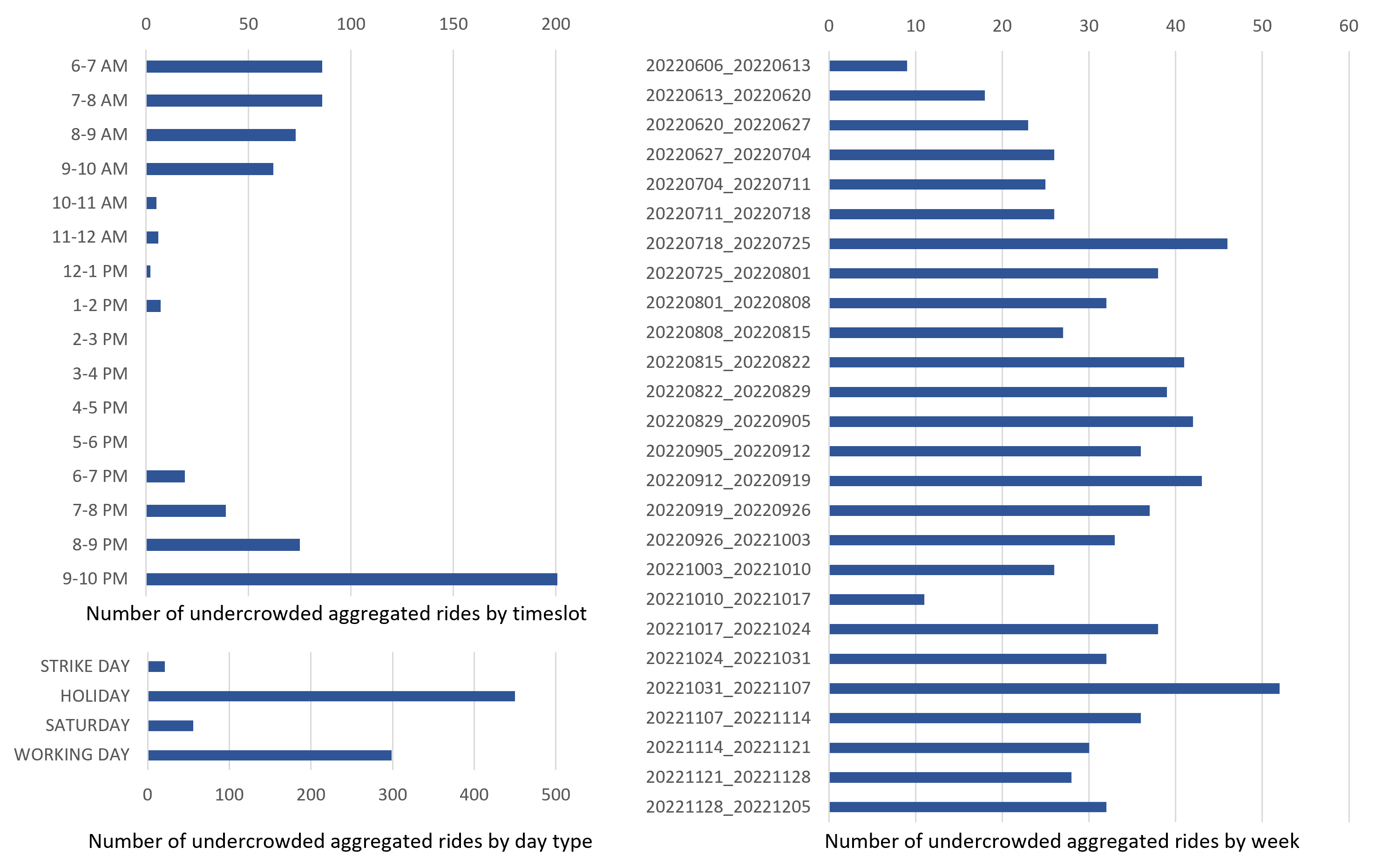}
      
      \caption{Fully undercrowded aggregated rides by time slot, day type, and month. Limit value for probability is set at $p\!=\!10$\%.}
      \label{fig:critical_ride}
\end{figure}

Using $p=0.1$ level, we examined the distribution of under-crowded rides across different time slots, day types, and months (refer to Figure \ref{fig:critical_ride}). Findings at the $p=0.1$ level indicate that fully undercrowded rides tend to manifest during the morning hours from 6 am to 9 am, as well as between 9 pm and 10 pm. Regarding day types, undercrowding situations are predominantly observed on holidays, with only a few instances occurring on working days. Analysis of occurrences by week does not reveal any discernible pattern, being the distribution of fully undercrowded rides almost homogeneous among weeks.

\section{CONCLUSIONS}\label{sec:conclusions}

This study aimed to contribute to the development of methodologies for the analysis of undercrowding and overcrowding of public transport means using novel sources of data. Leveraging the spatio-temporal granularity of data from APC systems, the study provides a methodology for investigating potential gaps between transportation offer and demand at the ride-segment level. In assessing the efficiency of the transportation service, the study focused on undercrowding situations, utilizing the Load Factor as an indicator of critical situations. A Generalized Linear Mixed Effects Model and a Generalized Mixed-Effect Random Forest were built to estimate the probability of undercrowding.

The model outputs streamline the recognition of under-crowded conditions across the entire journey, as well as critical issues confined to specific segments or subsections of the route. The integration of segment-level and ride-level analyses offers a holistic view, aiding informed decision-making for optimizing and sustaining public transportation systems. This approach allows for tailored interventions to address specific problems, whether they are concentrated in a particular segment or distributed throughout the entire journey, contributing to more effective problem-solving and operational efficiency. Specifically, the analysis of random effects can be interpreted to identify segments more likely associated with undercrowding situations. On the other hand, the analysis of fixed effects allows for the analysis of temporal and weather effects on the probability of having undercrowding, as well as analyzing the relationships between them (e.g. Time slot and Day type or Week and Day type). The analysis of ride-level findings can be employed to identify rides associated with a probability of undercrowding above the predefined limit for each segment of the entire route. Subsequently, through an examination of the temporal distribution, recurrent patterns can aid in identifying persistent gaps between supply and demand. This, in turn, suggests potential areas for improvement in transportation service planning. 

The study has practical implications for public transport operators, providing tools for analyzing the probability of undercrowding or overcrowding of public transport means from large-scale datasets. Notably, the model serves a dual purpose, functioning both as a monitoring system for critical situations and as a robust scenario analysis tool. Firstly, by facilitating the analysis of the spatio-temporal distribution of undercrowded rides while effectively isolating the influence of weather factors and specific event data, the model assists transport operators in promptly identifying critical situations that arise from the delicate balance between demand and supply. Secondly, the model can be used as a tool for scenario analysis, enabling the exploration of how changes in input variables impact its predictions. This capability allows for a comprehensive understanding of the probabilities associated with undercrowding and overcrowding under varying conditions. Such insights are invaluable for transportation decision-making and planning, particularly in the realm of tactical planning decisions. For instance, the model can contribute to defining strategies aimed at enhancing service levels and minimizing operational costs, including decisions related to frequency setting. Thus, the model's practical implications extend beyond monitoring, offering a data-driven tool for decision support in public transportation management.

However, the study has limitations that could guide future research. Despite the developed tools offering significant insights into recurrent undercrowding situations, the resulting insights must be complemented by other relevant information to make decisions on transportation network design and planning. Examples of these include constraints arising from resource allocation and management and minimum legal requirements on the public transport accessibility of a geographical area.

\vspace{3cm}
\textbf{Acknowledgements}:
The authors thank Azienda Trasporti Milanesi S.p.A. both for providing the data and for the indications of technical expertise on the problem of interest.

\textbf{Funding}: The authors acknowledge the support by MUR, grant Dipartimento di Eccellenza 2023-2027, Dipartimento di Matematica, Politecnico di Milano. Arianna Burzacchi's work has been further supported by the Next Generation EU Programme REACT-EU through the PON Ph.D. scholarship “Development of innovative Eulerian privacy-preserving data analysis tools for designing more sustainable and climate-friendly human mobility services and infrastructures from high-resolution location data”.


\bibliography{our}

\end{document}